\documentclass[conference]{IEEEtran}
\IEEEoverridecommandlockouts
\usepackage{cite}
\usepackage{amsmath,amssymb,amsfonts}
\usepackage{algorithmic}
\usepackage{graphicx}
\usepackage{textcomp}
\usepackage{xcolor}
\def\BibTeX{{\rm B\kern-.05em{\sc i\kern-.025em b}\kern-.08em
    T\kern-.1667em\lower.7ex\hbox{E}\kern-.125emX}} 
\usepackage{graphicx}
\usepackage{subcaption}
\usepackage{svg}
\usepackage{float}
\usepackage{booktabs}
\usepackage{tabularx}
\usepackage{svg}

\begin{document} 

%\tableofcontents

\title{Investigation of Feature Selection and Pooling Methods for Environmental Sound Classification\\
%\title{Investigation of Feature Selection and Pooling Methods for Efficient Environmental Sound Classification\\
%\title{Investigation of Feature Selection Methods for Efficient Environmental Sound Classification\\
}

\author{
%\IEEEauthorblockN{1\textsuperscript{st} Parinaz Binandeh Dehaghani*}
\IEEEauthorblockN{Parinaz Binandeh Dehaghani*}
\IEEEauthorblockA{\textit{SYSTEC-ARISE, Faculty of Engineering} \\
\textit{University of Porto, Portugal}\\
%Porto, Portugal \\
up202100618@edu.fe.up.pt}
*Corresponding author
~\\
\and
%\IEEEauthorblockN{2\textsuperscript{nd} Danilo Pena}
\IEEEauthorblockN{Danilo Pena}
\IEEEauthorblockA{\textit{ResoSight} \\
%\textit{University of Porto}\\
Montreal, Canada \\
danilo.pena@resosight.com}
~\\
\and
%\IEEEauthorblockN{3\textsuperscript{rd} A. Pedro Aguiar}
\IEEEauthorblockN{A. Pedro Aguiar}
\IEEEauthorblockA{\textit{SYSTEC-ARISE, Faculty of Engineering} \\
\textit{University of Porto, Portugal}\\
%Porto, Portugal \\
pedro.aguiar@fe.up.pt}
% ~\\
% \and

% \and

% \and
}

\maketitle

\begin{abstract}
This paper explores the impact of dimensionality reduction and pooling methods for Environmental Sound Classification (ESC) using lightweight CNNs. We evaluate Sparse Salient Region Pooling (SSRP) and its variants, SSRP-Basic (SSRP-B) and SSRP-Top-K (SSRP-T), under various hyperparameter settings and compare them with Principal Component Analysis (PCA). Experiments on the ESC-50 dataset demonstrate that SSRP-T achieves up to 80.69\% accuracy, significantly outperforming both the baseline CNN (66.75\%) and the PCA-reduced model (37.60\%). Our findings confirm that a well-tuned sparse pooling strategy provides a robust, efficient, and high-performing solution for ESC tasks, particularly in resource-constrained scenarios where balancing accuracy and computational cost is crucial.
\end{abstract}


\section{Introduction}

Environmental Sound Classification (ESC) has emerged as a relevant research area within audio signal processing, driven by its many practical applications in areas such as smart cities, audio surveillance systems, healthcare monitoring, security systems, and Internet of Things (IoT) deployments. ESC involves the automated recognition and categorization of audio events occurring in various environmental contexts, such as urban, domestic, natural, and human-related sounds. Unlike speech or music recognition, ESC faces distinct challenges due to the inherent variability, diverse frequency content, low signal-to-noise ratio, and complex temporal structures characteristic of environmental audio data. Moreover, advances in ESC often inspire customized sound classification solutions in industry, enabling novel applications that were previously infeasible due to technical limitations~\cite{bansal2022environmental}. 
Recent advances in ESC highlight the effectiveness of attention and transformer architectures. The Audio Spectrogram Transformer (AST) adapts Vision Transformers to log-mel spectrograms, achieving strong results on ESC-50 by leveraging global context with multi-head self-attention~\cite{gong2021ast}. Variants like PaSST (Patchout Spectrogram Transformer) further improve efficiency by reducing computation via patch dropping while maintaining high accuracy~\cite{koutini2021efficient}. An important component for achieving effective audio classification performance is the use of appropriate and efficient feature selection techniques. 
%Early studies presented Principal Component Analysis (PCA) for feature selection with a set of features, enhancing the discriminative power of an SVM model~\cite{rabaoui2008using}. Multiple machine learning classifiers were evaluated for three sets of features using PCA, which allowed the selection of the set with the highest classification accuracy~\cite{bountourakis2015machine}. For music genre classification, feature selection was explored, inspired by data mining techniques, identifying the most informative features from a broader set, and selecting them for performance improvements of the classifier~\cite{mitra2014efficient}.
Several studies have employed Principal Component Analysis (PCA) for feature selection to improve model performance. For instance, PCA has been used to enhance the discriminative capability of an SVM model by selecting the most relevant features~\cite{rabaoui2008using}. In another study, multiple machine learning classifiers were evaluated across three different feature sets, where PCA facilitated the identification of the set yielding the highest classification accuracy~\cite{bountourakis2015machine}.
Recent progress in Deep Learning has greatly enhanced the performance of ESC systems, with Convolutional Neural Networks (CNNs) playing a key role, as they can not only achieve high classification accuracy but also extract meaningful features for broader tasks. However, these improvements typically come at the cost of increased model complexity and computational demands, limiting their applicability to resource-constrained environments such as IoT devices and embedded systems~\cite{fang2022fast}. Consequently, efficient and robust feature selection methods are vital to addressing this issue, as they can substantially reduce network complexity, accelerate training and inference processes, and enhance the CNN’s capacity to learn meaningful representations. Traditional feature selection methods have been widely employed to reduce dimensionality and enhance classifier efficiency. While these methods help mitigate redundancy and improve computational efficiency, they often suffer from certain limitations when applied to complex time-frequency audio representations~\cite{ren2025group}. These techniques are primarily designed to rank or project features based on statistical measures, lacking the capacity to capture intricate temporal dependencies and localized discriminative regions effectively. Furthermore, their static nature makes it challenging to adaptively focus on the most relevant sound events, particularly in the presence of noise and overlapping acoustic patterns.

Another dimensionality reduction technique often used in the context of neural networks is the pooling method. Which primarily downsamples feature maps, focusing the model on important local features~\cite{zafar2022comparison}. Similarly to feature selection techniques, it helps to reduce overfitting and reduce dimensionality, consequently reducing computational efforts. Recently, state-of-the-art methods, such as Sparse Salient Region Pooling (SSRP), have been introduced for ESC~\cite{seresht2022environmental}. SSRP leverages the sparse and salient regions within time-frequency audio representations, allowing CNNs to prioritize learning from the most discriminative audio patterns. By focusing on the most critical temporal and frequency regions, SSRP not only enhances model interpretability but also improves classification accuracy in ESC tasks. This makes it a promising alternative to traditional selection methods, addressing their shortcomings by dynamically filtering irrelevant features and amplifying significant ones. Nevertheless, a thorough comparative analysis between the SSRP method and conventional feature selection techniques remains crucial to fully understand their relative benefits and limitations.
There is limited work that systematically compares explicit sparsity-aware pooling against classical dimensionality reduction under a controlled CNN setup for ESC. The SSRP family was originally shown to outperform GAP and other pooling choices within lightweight CNNs for ESC-50 \cite{seresht2022environmental}, but direct head-to-head comparisons with PCA-based pipelines have not been reported to our knowledge.
This paper investigates the efficacy of PCA compared to the SSRP technique, using lightweight CNNs in the context of ESC. Through this comparative analysis, it aims to identify optimal feature selection practices that balance computational efficiency with classification performance, thereby guiding future developments and applications in environmental audio classification.

This paper is organized as follows: Section II introduces the dimensionality reduction using PCA; Section III details the pooling models; Section IV describes the experimental setup; Section V presents the results and their analysis; and Section VI offers concluding remarks and outlines potential directions for future work.

\section{Dimensionality Reduction using PCA}

PCA is a dimensionality reduction technique that transforms high-dimensional data into a set of orthogonal components, each capturing a portion of the variance present in the original dataset. By selecting a subset of these principal components, PCA effectively reduces feature dimensionality while preserving the most informative aspects of the data. In the context of ESC, PCA can potentially improve computational efficiency and classification accuracy by removing redundant and correlated features. By projecting audio features onto a lower-dimensional space, PCA facilitates more robust and generalized classifier performance, making it particularly beneficial for applications with limited computational resources \cite{mitra2014efficient}. Let $\mathbf{X} \in \mathbb{R}^{n \times d}$ be a dataset composed of $n$ samples and $d$ features, where each row of $\mathbf{X}$ corresponds to a feature vector. PCA performs a linear transformation:

\begin{equation}
\mathbf{Z} = \mathbf{X} \mathbf{W} \text{,}
\label{eq:pca_projection}
\end{equation}

\noindent where $\mathbf{Z} \in \mathbb{R}^{n \times k}$ is the transformed dataset in the reduced $k$-dimensional space ($k < d$), and $\mathbf{W} \in \mathbb{R}^{d \times k}$ is a projection matrix whose columns are the top $k$ eigenvectors of the covariance matrix $\mathbf{\Sigma}$ of $\mathbf{X}$. The covariance matrix is defined as:

\begin{equation}
\mathbf{\Sigma} = \frac{1}{n - 1} \mathbf{X}^\top \mathbf{X} \text{.}
\label{eq:covariance_matrix}
\end{equation}

PCA solves the following eigenvalue problem:

\begin{equation}
\mathbf{\Sigma} \mathbf{w}_i = \lambda_i \mathbf{w}_i \quad \text{for } i = 1, \dots, d \text{,}
\label{eq:eigen_decomposition}
\end{equation}

\noindent where $\lambda_i$ are the eigenvalues indicating the variance explained by each component, and $\mathbf{w}_i$ are the corresponding eigenvectors, sorted in decreasing order of $\lambda_i$. By selecting the top $k$ eigenvectors associated with the largest eigenvalues, PCA retains the components that explain the most variance in the data while discarding those contributing the least.

\section{Pooling Models}
% GAP: talk about global averaging pooling
\subsection{Traditional Pooling Methods} 

Traditional pooling methods, such as \textit{Max Pooling} and \textit{Average Pooling}, are fundamental techniques in CNNs for dimensionality reduction and feature aggregation. Max Pooling operates by selecting the maximum activation value from each local region of the feature map, effectively capturing the most prominent features while discarding weaker activations. This strategy enhances feature robustness and reduces spatial variance, making it well-suited for capturing distinct patterns. On the other hand, Average Pooling computes the mean value of each local region, preserving more generalized information across spatial dimensions. While both methods contribute to reducing computational complexity and mitigating overfitting, they often fail to emphasize the most discriminative regions, particularly in complex auditory scenes like ESC. In these scenarios, the ability to selectively prioritize salient audio patterns is crucial, motivating the exploration of advanced pooling strategies such as the SSRP.

\subsection{Sparse Salient Region Pooling (SSRP) Methods}

Advanced pooling methods like SSRP, initially developed for image processing and computer vision tasks, have been adapted to enhance CNN for ESC systems by focusing on sparse and salient regions within time-frequency audio representations \cite{hou2011image}. SSRP applied to ESC focuses on selectively capturing sparse and salient regions within the time-frequency audio representations, which contain the most informative and discriminative characteristics. Unlike traditional pooling methods, such as Global Average Pooling (GAP), SSRP does not treat all regions equally but rather prioritizes sparse and highly relevant regions \cite{seresht2022environmental}. This selective approach significantly enhances the CNN's ability to learn meaningful patterns from the data, improving classification performance while simultaneously reducing computational requirements. SSRP thus addresses key limitations associated with conventional pooling strategies, making it particularly suitable for ESC applications in resource-constrained environments.

In this work, we focus on two SSRP pooling strategies~\cite{seresht2022environmental}. They are: Sparse Salient Region Pooling - Basic (SSRP-B), Top-K Salient Region Pooling (SSRP-T), discussed in detail below.

\subsubsection{SSRP-B}

SSRP-B is the simplest variant of SSRP, where a fixed-size temporal window \( W \) is used to pool feature activations. It selects only the most salient region, in this case, the highest mean activation, within each window, effectively filtering out less informative regions.

The pooled representation is computed as follows:

\begin{equation}
    z_c(f) = \frac{1}{W} \sum_{i=1}^{W} m_c^D(t + i, f) \text{,}
\end{equation}

\noindent where \( m_c^D(t, f) \) represents the activation value for channel \( c \) at time \( t \) and frequency \( f \), \( W \) is the window size.

\subsubsection{SSRP-T}

SSRP-T extends the basic pooling mechanism by allowing the model to consider multiple high-activation regions instead of just the single most active window. Here, the top \( K \) activations, sorted by their mean values, are averaged to form the final pooled representation, capturing more salient regions distributed over time.
The pooled representation is computed as:

\begin{equation}
    z_c(f) = \frac{1}{K} \sum_{k=1}^{K} s_c^{[k]}(f) \text{,}
\end{equation}

\noindent where \( K \) is the number of top activations selected, \( s_c^{[k]}(f) \) denotes the \( k^{th} \) highest mean activation for channel \( c \) and frequency \( f \).

\section{Experimental Setup}

\subsection{Dataset}

The experiments in this study were conducted on the ESC-50 dataset \cite{piczak2015esc}, a publicly available and widely used benchmark for environmental sound classification. The dataset consists of 2000 audio recordings of 5 seconds each, sampled at 44.1 kHz, and distributed equally across 50 sound classes. These classes are grouped into five major categories: animals, natural soundscapes, human non-speech sounds, interior/domestic sounds, and exterior/urban noises. All sound clips are provided in mono-channel WAV format and are pre-organized with metadata including the filename, category, fold number, and class label. For this study, we followed the standard stratified 5-fold cross-validation protocol to evaluate performance fairly and consistently across all categories. 
\subsection{CNN Architecture} 

\begin{figure}[ht]
    \centering
    \includegraphics[width=\linewidth]{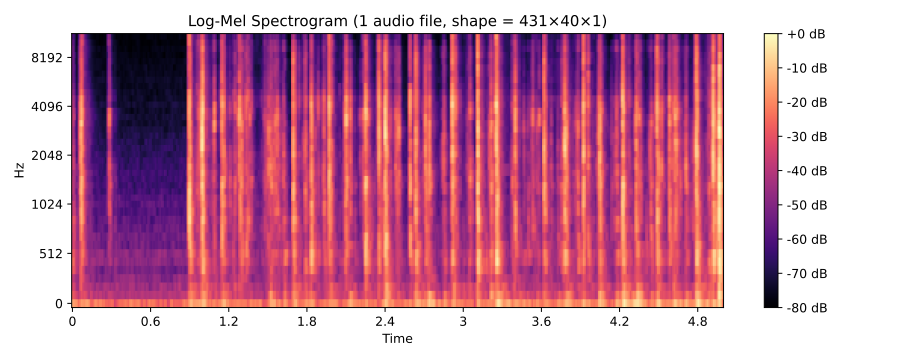}
    \caption{Log-Mel spectrogram representation of an audio sample used as input to the CNN. The spectrogram has a shape of (431, 40, 1), corresponding to 431 time frames and 40 mel frequency bins. The color intensity indicates signal power in decibels (dB) across time and frequency.}
    \label{fig:Mel_Example}
\end{figure}

All models in this study share a common CNN backbone architecture, designed for efficiency and effectiveness in environmental sound classification. The architecture is inspired by lightweight CNNs\cite{seresht2022environmental} used in audio tasks, with the only difference between models being the choice of global pooling strategy (SSRP-B or SSRP-T). This design ensures a fair comparison across pooling methods, isolating the effect of sparsity mechanisms on model performance. The input to the network is a log-mel spectrogram with shape (431,40,1), where 431 represents the number of time frames and 40 denotes the mel frequency bins. An example of such an input is illustrated in Figure \ref{fig:Mel_Example}.

\begin{figure}
    \centering
    \includegraphics[width=1\linewidth]{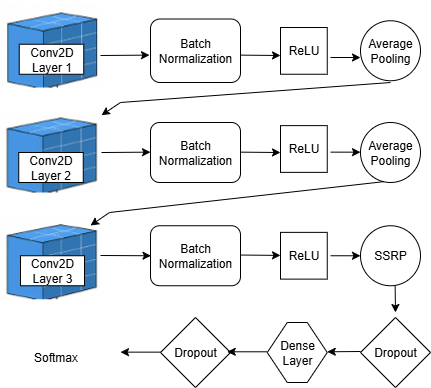}
    \caption{Common CNN Backbone Architecture}
    \label{fig:CNN_Model}
\end{figure}

These spectrograms are normalized and padded/truncated to maintain consistent input dimensions across all samples. The overall CNN architecture is illustrated in Figure~\ref{fig:CNN_Model}. The model consists of three convolutional layers with increasing filter sizes: Conv2D Layer 1 with 32 filters, kernel size 3×3. A Conv2D Layer 2 with 64 filters, kernel size 3×3. And a Conv2D Layer 3 with 128 filters, kernel size 3×3. Each convolutional layer is followed by Batch Normalization to stabilize learning by normalizing the activations, ReLU activation to introduce non-linearity, and Average Pooling (after the first and second layers) with a pool size of 2×2, reducing spatial dimensions while retaining important feature information. After the final convolutional layer, an SSRP layer is applied. Depending on the experiment, this is either SSRP-B or SSRP-T, each controlling sparsity differently. This is the only part of the architecture that varies across models to evaluate the impact of the pooling strategy.

The output of the SSRP layer is flattened and passed through a Dense layer with 128 units and ReLU activation, providing high-level feature representation, followed by a Dropout layer with a rate of 0.5, helping to prevent overfitting by randomly deactivating neurons during training.
Finally, the output layer is a Dense layer with 50 units (one for each ESC-50 class) and softmax activation, providing class probabilities.
For training, we employed the Mixup augmentation technique with an alpha value of 0.2, which improves generalization by linearly interpolating pairs of training examples and their labels. The models were trained with a batch size of 64 for 700 epochs, using stochastic gradient descent (SGD) with a learning rate of 0.05 and a momentum of 0.9. These hyperparameter choices were kept consistent across all pooling variants to ensure fair comparison.

\subsection{Implementation Details}

All experiments were conducted on Google Colab, utilizing its provided NVIDIA GPU for accelerated computation. The implementation was carried out using the Keras deep learning API with a TensorFlow backend for model training and evaluation. Feature extraction from audio files was performed with the Librosa library.

\section{Results and Analysis}

\subsection{Effect of PCA on CNN}

\begin{figure}[t!]
    \centering
    \includegraphics[width=\linewidth]{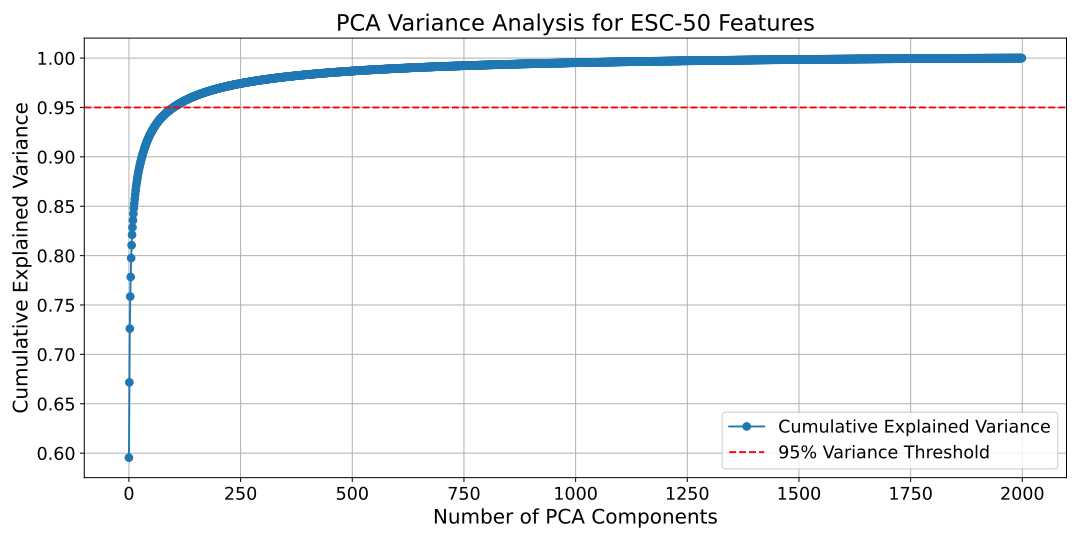}
    \caption{Cumulative explained variance curve for PCA applied to the ESC-50 log-Mel spectrogram features. The red dashed line indicates the 95\% variance threshold, achieved with 101 components.}
    \label{fig:PCA_variance}
\end{figure}

To explore the impact of dimensionality reduction on environmental sound classification, Principal Component Analysis (PCA) was applied to log-Mel spectrograms derived from the ESC-50 dataset. Each 5-second audio clip was converted into a 40 × 428 log-Mel spectrogram and flattened into a 17,120-dimensional vector. After standardization, PCA was used to retain 95\% of the total variance, resulting in 101 principal components and achieving a 99.41\% reduction in dimensionality. As shown in Figure \ref{fig:PCA_variance}, the curve rises sharply, indicating that most of the data's variance is captured by the first few principal components. The reduced features were reshaped to fit the input format of a lightweight CNN, which consisted of three convolutional blocks, batch normalization, ReLU activation, average pooling, and a final dense layer. Training was conducted with 5-fold cross-validation and Mixup data augmentation to enhance generalization and robustness.

Despite the computational and memory efficiency gained through PCA, the classification performance significantly declined. The CNN trained on the full log-Mel spectrogram (no dimensionality reduction) achieved an accuracy of 66.75\%, while the CNN trained on PCA-reduced features achieved only 37.60\%. This demonstrates that although PCA effectively compresses data, it may discard subtle yet discriminative features essential for classification, particularly when using models like CNNs that are designed to exploit local time-frequency structures in spectrograms.

%\begin{figure}[ht]

\subsection{Analysis of SSRP Variants}

\subsubsection{Analysis of SSRP-B with Varying Window Sizes} 
\begin{figure*}[ht]
    \centering
    \begin{subfigure}[t]{0.48\textwidth}
        \centering
        \includegraphics[width=\linewidth]{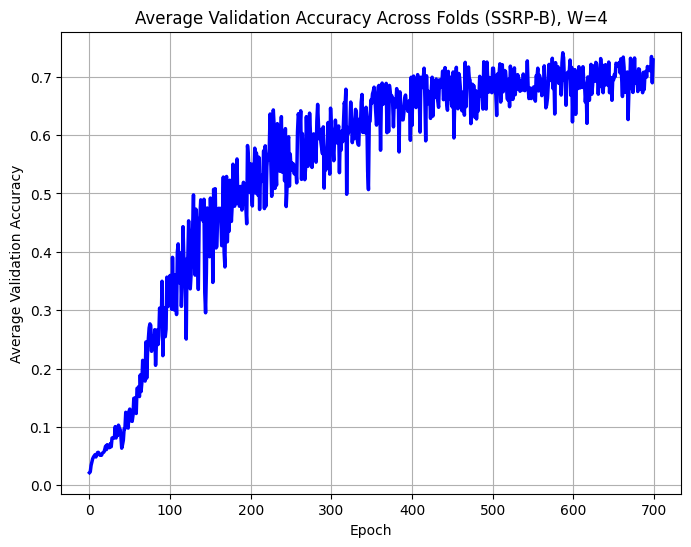}
        \caption{Validation accuracy (SSRP-B, $W=4$)}
        \label{fig:ssrpB-w4}
    \end{subfigure}
    \hfill
    \begin{subfigure}[t]{0.48\textwidth}
        \centering
        \includegraphics[width=\linewidth]{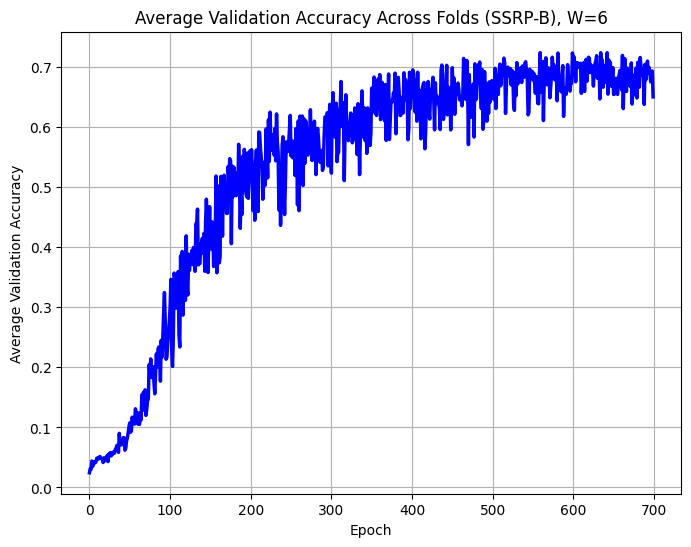}
        \caption{Validation accuracy (SSRP-B, $W=6$)}
        \label{fig:ssrpB-w6}
    \end{subfigure}
    \caption{Comparison of validation accuracy for different window sizes in SSRP-B pooling. (a) Validation accuracy for SSRP-B with $W=4$. (b) Validation accuracy for SSRP-B with $W=6$.}
    \label{fig:ssrpB-comparison}
\end{figure*}

The pooling operation in SSRP-B selects only the most salient region—specifically, the highest mean activation—within each window, thereby introducing sparsity in the representation. To evaluate the impact of the window size W, we tested several values, as shown in table \ref{tab:SSRP-hyperparam}. When $W=4$, the model achieved its highest validation accuracy. %Figure \ref{fig:SSRPB_W4} shows the confusion matrix for SSRP-B with $W=4$.
Smaller window sizes allow the model to focus more precisely on short-term, high-importance sound events. This is particularly beneficial in environmental sound classification, where transient sounds (e.g., dog barks, door knocks) play a crucial role. In contrast, larger window sizes, such as $W=6$, aggregate a broader temporal context. While this may help in detecting long-duration events, it tends to dilute the effect of sharp, transient cues, leading to reduced accuracy.
These results suggest that more aggressive sparsity (i.e., smaller W) can be advantageous, as it helps filter out less informative temporal regions. Although the optimal window size in SSRP-B may be task-dependent, our experiments indicate that smaller windows are generally more effective. This highlights the importance of considering the temporal characteristics of sound classes when tuning W. As illustrated in Figure \ref{fig:ssrpB-comparison}, the validation accuracy curves demonstrate that the model with $W=4$ consistently outperforms the configuration with $W=6$, particularly during the early training epochs, underscoring the effectiveness of more aggressive temporal sparsity in SSRP-B.

%\begin{table}[ht]
\begin{table}[t!]
\centering
\caption{Comparison of CNN + SSRP-B and CNN + SSRP-T models with different hyperparameters}
\begin{tabular}{|l|c|c|c|}
\hline
\textbf{Model} & \textbf{Input} & \textbf{Hyper Param} & \textbf{Accuracy (\%)} \\
\hline
CNN + SSRP-B & Log Mel & $W=2$  & 71.15 \\
CNN + SSRP-B & Log Mel & $W=4$  & 72.85 \\
CNN + SSRP-B & Log Mel & $W=6$  & 65.05 \\
CNN + SSRP-B & Log Mel & $W=8$  & 66.09 \\
CNN + SSRP-T & Log Mel & $K=4$  & 75.20 \\
CNN + SSRP-T & Log Mel & $K=8$  & 77.60 \\
CNN + SSRP-T & Log Mel & $K=10$ & 80.60 \\
CNN + SSRP-T & Log Mel & $K=12$ & \textbf{80.69} \\
CNN + SSRP-T & Log Mel & $K=14$ & 78.65 \\
CNN + SSRP-T & Log Mel & $K=16$ & 70.59 \\
\hline
\end{tabular}
\label{tab:SSRP-hyperparam}
\end{table}

%%%%%%%%%%%%%%%%%%%%%%%%%%
% --- Confusion Matrices (Full Column Width & Larger Font) ---
%%%%%%%%%%%%%%%%%%%%%%%%%%

%\begin{figure}%[H]
 %   \centering
  %  \includegraphics[width=\columnwidth]{figures/SSRPB_W4.png}
   % \caption{Confusion matrix for SSRP-B with $W=4$.}
    %\label{fig:SSRPB_W4}
%\end{figure} 

%\begin{figure}[H]
    %\centering
    %\includegraphics[width=\columnwidth]{figures/SSRPB_W6.png}
    %\caption{Confusion matrix for SSRP-B with $W=6$.}
    %\label{fig:SSRPB_W6}
%\end{figure}

%%%%%%%%%%%%%%%%%%%%%%%%%%
% --- Validation Accuracy Comparison (Full Width Subfigures) ---
%%%%%%%%%%%%%%%%%%%%%%%%%%

\subsubsection{Analysis of SSRP-T with Varying Top-K Values}

To study the effect of \( K \), we tested a range of values including \( K = 4, 8, 10, 12 \) and further examined the impact beyond this range. As \( K \) increases from 4 to 12, validation accuracy consistently improves. This indicates that aggregating more salient regions allows the model to build a richer representation of the input, capturing a wider variety of temporal patterns relevant for classification. However, when \( K > 12 \), the model starts to include less informative or even noisy temporal intervals. This results in a decline in accuracy, as the pooling layer loses its selectivity and becomes overly inclusive. Thus, there exists an optimal range for \( K \), where the model benefits from multi-region aggregation without overfitting to irrelevant features. SSRP-T balances sparsity and richness in representation better than SSRP-B, especially when sound events exhibit complex or repeated structures over time. 
The impact of different values of K in the SSRP-T layer on validation accuracy is illustrated in table \ref{tab:SSRP-hyperparam}. As shown, increasing K from 4 to 12 results in an improvement in average validation accuracy.

This improvement is attributed to the ability of larger window sizes to better capture long-term dependencies and evolving sound events that are commonly present in environmental sounds. For instance, sounds such as engine noise, rainfall, or wind exhibit long-lasting temporal patterns that require broader context for accurate representation. The larger multi-scale configuration enables the model to aggregate these temporal patterns more effectively, reducing fragmentation and preserving important sequential information. In contrast, the smaller-scale setting [2, 4, 6] tends to fragment long-duration events across multiple windows, leading to the loss of critical temporal cues. Consequently, the [6, 10, 14] configuration proves to be more effective at representing both short-term transients and long-term events, enhancing classification accuracy by addressing the limitations of single-scale pooling.

%%%%%%%%%%%%%%%%%%%%%%%%%%%%%%%%% 
\begin{figure*}[ht]
    \centering
    \begin{subfigure}[t]{0.48\textwidth}
        \centering
        \includegraphics[width=\linewidth]{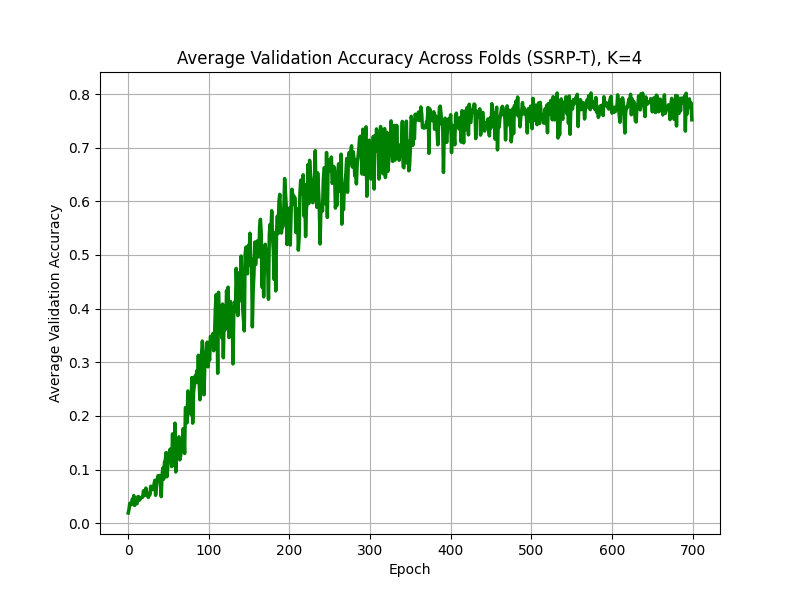}
        \caption{Validation accuracy (SSRP-T, $K=4$)}
        \label{fig:ssrpT-K4}
    \end{subfigure}
    \hfill
    \begin{subfigure}[t]{0.48\textwidth}
        \centering
        \includegraphics[width=\linewidth]{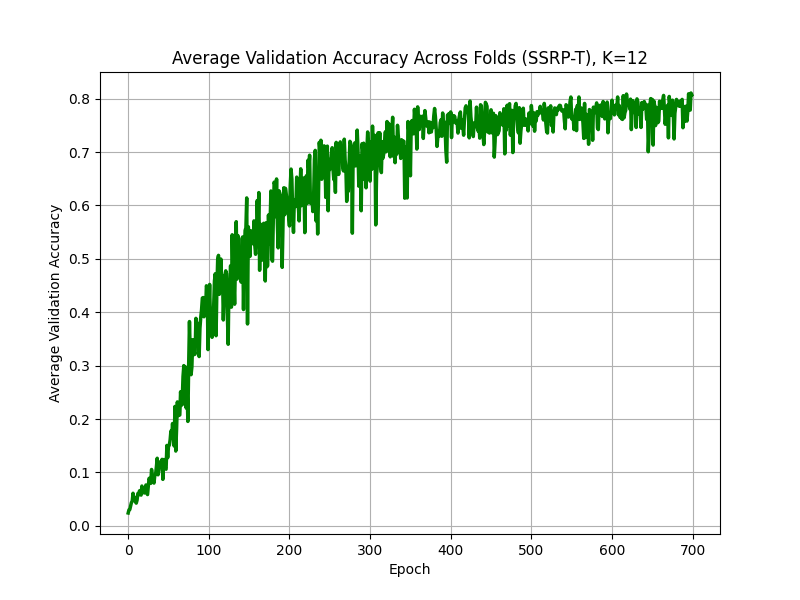}
        \caption{Validation accuracy (SSRP-T, $K=12$)}
        \label{fig:ssrpT-K12}
    \end{subfigure}
    \caption{Comparison of validation accuracy for different values of K in SSRP-T pooling. (a) Validation accuracy for SSRP-T with $K=4$. (b) Validation accuracy for SSRP-T with $K=12$.}
    \label{fig:ssrpT-comparison}
\end{figure*}
%%%%%%%%%%%%%%%%%%%%%%%%%%%%%%%%%%%%%% 

\subsection{Comparison Between PCA and SSRP on Accuracy}

A comparative analysis was conducted to evaluate the classification performance of PCA-based dimensionality reduction against the proposed SSRP mechanism integrated with CNNs. Applying PCA to reduce the input dimension from 17,120 to 101 components (99.41\% reduction) resulted in a significant drop in classification accuracy—from 66.75\% (no reduction) to 37.60\%. This highlights that while PCA effectively compresses feature space, it discards critical local patterns essential for accurate sound classification. In contrast, the SSRP-enhanced models, which preserve spatial structure and introduce sparsity through selective pooling, consistently outperformed the PCA-based model. Specifically, the CNN+SSRP-B model achieved up to 72.85\% accuracy at $W=4$, and CNN+SSRP-T peaked at 80.69\% accuracy at $K=12$, both surpassing the baseline CNN and the PCA-reduced variant. These results clearly demonstrate that SSRP not only avoids the detrimental loss of discriminative information seen in PCA but also enhances model performance by focusing on salient temporal regions. Therefore, SSRP is a more effective strategy for feature aggregation in environmental sound classification tasks.

\begin{table}[H]
\centering
\caption{Comparison of pooling strategies using Log Mel with hyperparameters, accuracy, and model complexity.}

\begin{tabular}{|c|c|c|c|c|}
    \hline
    \textbf{Pooling} & \textbf{\# of Layers} & \textbf{Hyper Param} & \textbf{Acc (\%)} & \textbf{\# of Params} \\ 
    \hline
    SSRP-B & 3 & $W=4$ & 72.85 & 527 K \\ 
    %\hline
    SSRP-T & 3 & $K=12$ & 80.69 & 527 K \\ 
    %\hline
    %SSRP-MS & 3 & $w=[4,6,8]$ & 43 & 527 K \\ 
    Baseline & 3 &   $-$   & 66.75 & 245 K \\
    %\hline
    %PCA\_MaxPooling & Log Mel & 3 & * & 53.25 & 182 K \\ 
    \hline
%\end{tabular}
\end{tabular}
\label{tab:pooling-comparison}
\end{table}

\section{Conclusion}

In this paper, we investigated the effectiveness of sparse pooling strategies and PCA-based dimensionality reduction for environmental sound classification. Using the ESC-50 dataset, we evaluated two SSRP variants—SSRP-B and SSRP-T—within a lightweight CNN architecture. As shown in Table~\ref{tab:pooling-comparison}, our results demonstrated that optimal hyperparameter settings significantly impact performance: a window size of $W=4$ for SSRP-B and a Top-K value of $K=12$ for SSRP-T yielded the best results, with SSRP-T achieving an accuracy of 80.69\%, and SSRP-B reaching 72.85\%. Both configurations significantly outperformed the baseline CNN (66.75\%) and the PCA-reduced model (37.60\%).

Notably, although the PCA approach resulted in a drastic 99.41\% reduction in feature dimensionality, it led to a considerable drop in classification performance. In contrast, the SSRP-based models enhanced accuracy while maintaining the same number of layers (3) and a comparable parameter count (527K) as the CNN+SSRP baseline. Interestingly, even the baseline model without sparse pooling had fewer parameters (245K) but delivered lower accuracy, suggesting that the parameter increase in SSRP-based models is justified by the performance gain. Overall, our study highlights the advantage of incorporating task-aware sparse pooling strategies into CNNs for environmental sound classification. These methods not only preserve essential temporal structures but also improve generalization and robustness while maintaining lightweight model complexity.

Future work will explore integrating SSRP with attention mechanisms, adaptive Top-K selection, and transformer-based architectures to further enhance temporal modeling and performance.

\section*{Acknowledgment}
This work was financially supported by SYSTEC – UID/00147,funded by FCT/MECI through national funds.It was also supported by ARISE–LA/P/0112/2020, funded by FCT/MECI through national funds, and by the project “F2ForAll – Development of a sustainable ‘From Farm to Fork’ strategy in the Mediterranean, integrating Optimal Control, Deep Learning, and Advanced Econometrics” – operation NORTE2030-FEDER-02700700, NORTE2030 under the SACCCT – Integrated R\&D Projects.

\bibliographystyle{IEEEtran}  % Citation style (unsrt, plain, alpha, etc.)
\bibliography{main}  % Name of the .bib file (without .bib extension)

\end{document}